\documentclass{article}

\usepackage[nonatbib, preprint]{neurips_2020}

\usepackage[utf8]{inputenc} 
\usepackage[T1]{fontenc}    
\usepackage{hyperref}       
\usepackage{url}            
\usepackage{booktabs}       
\usepackage{amsfonts}       
\usepackage{nicefrac}       
\usepackage{microtype}      
\usepackage{graphicx}       
\usepackage{color, colortbl} 
\usepackage{multirow} 
\usepackage{longtable}
\usepackage[most]{tcolorbox} 
\usepackage{lipsum} 
\usepackage[utf8]{inputenc}
\usepackage{tcolorbox}
\usepackage{amsmath}
\usepackage{tcolorbox}

\usepackage{booktabs} 
\usepackage{tabularx} 
\usepackage{geometry} 

\usepackage[utf8]{inputenc}
\usepackage{enumitem}

\definecolor{headercolor}{gray}{0.85}
\definecolor{maxvalue}{rgb}{0.92,1,0.92} 

\tcbuselibrary{skins}

\title{Enhancing Guardrails for Safe and Secure Healthcare AI}

\author{
  Ananya Gangavarapu\\
  AI Engineer, 
  Ethriva Inc.\\
  \texttt{ananya@ethriva.ai} \\
}

\begin{document}

\maketitle

\begin{abstract}

 Generative AI holds immense promise in addressing global healthcare access challenges, with numerous innovative applications now ready for use across various healthcare domains. However, a significant barrier to the widespread adoption of these domain-specific AI solutions is the lack of robust safety mechanisms to effectively manage issues such as hallucination, misinformation, and ensuring truthfulness. Left unchecked, these risks can compromise patient safety and erode trust in healthcare AI systems. While general-purpose frameworks like Llama Guard are useful for filtering toxicity and harmful content, they do not fully address the stringent requirements for truthfulness and safety in healthcare contexts. This paper examines the unique safety and security challenges inherent to healthcare AI, particularly the risk of hallucinations, the spread of misinformation, and the need for factual accuracy in clinical settings. I propose enhancements to existing guardrails frameworks, such as Nvidia NeMo Guardrails, to better suit healthcare-specific needs. By strengthening these safeguards, I aim to ensure the secure, reliable, and accurate use of AI in healthcare, mitigating misinformation risks and improving patient safety.

\end{abstract}

\section{Introduction}

Generative AI, particularly Large Language Models (LLMs), has quickly become a transformative force in global healthcare, offering a wide range of applications that enhance clinical operations. These models are used in various contexts, from assisting in tasks like summarizing medical records \cite{stanfordHAI2023, tang2023evaluating}and research papers to playing pivotal roles in complex scenarios, such as automating diagnostics in Intensive Care Units (ICUs) and analyzing fast-changing electronic health records (EHRs). The true power of LLMs lies in their ability to process vast volumes of medical data, synthesize it into meaningful insights, and generate highly contextual, relevant information. For instance, they can help identify subtle patterns in patient histories, recommend personalized treatment plans based on the latest medical literature, and provide real-time decision support in emergency settings. This capability not only improves healthcare delivery worldwide but also enables more efficient decision-making and streamlines clinical workflows—an essential benefit as healthcare systems grapple with increasing complexity and resource constraints.

However, despite the potential, LLMs also introduce significant risks that could undermine their effectiveness in critical care environments. Issues like hallucinations\cite{McKenna2023}—where the model generates false or misleading information—can be particularly dangerous in medical settings. A hallucinated diagnosis or treatment recommendation in an ICU could have life-threatening consequences. Similarly, jailbreaking—manipulating the model \cite{Wu2024}to bypass built-in safeguards—poses another serious concern, potentially allowing the generation of harmful or inappropriate responses. Misinformation is another pressing issue, as inaccurate information on conditions, treatments, or drug interactions could lead to harmful medical decisions.

While some safety concerns, like hallucinations, are being addressed through strategies such as integrating external knowledge bases for real-time fact-checking \cite{Yan2024} and refining prompting techniques, these approaches have yet to be rigorously validated for healthcare-specific LLMs. For example, linking an LLM to trusted medical databases like PubMed \cite{PubMed} or UpToDate \cite{UpToDate}can improve its ability to cross-reference and verify clinical information, but the effectiveness of these methods in preventing hallucinations in high-stakes environments, like emergency care or oncology, remains largely unexplored.

 To address these concerns, innovative frameworks such as NVIDIA NeMo Guardrails \cite{Rebedea2023} and Llama Guard \cite{Meta2023} have emerged as promising solutions. NeMo Guardrails offers customizable "rails" for content filtering, allowing developers to create healthcare-specific filters that account for the complexities of medical terminology and ensure that critical discussions are accurate and safe. Llama Guard, meanwhile, focuses on strengthening the security of LLMs by preventing jailbreaking and unauthorized manipulations—vital for maintaining the integrity and trustworthiness of medical AI systems.
By integrating NeMo Guardrails with Llama Guard’s security mechanisms, a more comprehensive framework could potentially address the unique risks associated with healthcare AI. This approach could facilitate real-time fact-checking against trusted medical knowledge bases while strengthening security protocols to minimize the risk of misuse. Together, these frameworks have the potential to mitigate issues such as hallucinations and misinformation, helping healthcare LLMs deliver more reliable and contextually accurate information.

This paper presents an approach for fusing NeMo Guardrails and Llama Guard, evaluating the effectiveness of this integrated approach using several medical datasets. Through these evaluations, the paper assesses how well the combined guardrails can enhance patient safety and maintain the integrity of healthcare systems, with the ultimate goal of supporting the broader adoption of AI in healthcare by ensuring the accuracy and reliability of information.

\section{Background}

In this section, I summarize the key issues associated with medical LLMs, focusing on their vulnerabilities and potential risks in healthcare applications. These models, while offering transformative potential, face challenges related to the accuracy and reliability of the information they generate. Among these challenges are issues such as information inaccuracy, system vulnerabilities, and the potential for misuse. In particular, LLMs can produce incorrect or irrelevant content, leading to concerns about the safety of their deployment in medical contexts. To address these issues systematically, I will explore the vulnerabilities of medical LLMs through a structured approach, covering hallucinations, jailbreaking, misinformation, and semantic errors.

\subsection{Jailbreaking and Security Vulnerabilities}

Medical LLMs are susceptible to jailbreaking, where the model’s built-in safeguards are bypassed, enabling the generation of harmful, inappropriate, or confidential content. This vulnerability severely undermines the trustworthiness and safety of medical AI, particularly in high-stakes healthcare settings where patient data, diagnostic information, and treatment recommendations must remain secure and protected. Jailbreaking can expose systems to significant risks, including breaches of confidentiality, the manipulation of clinical decision-making processes, and the propagation of dangerous misinformation that may harm both patients and healthcare professionals.

The consequences of jailbreaking are especially critical in medical contexts where the ethical use of data, reliability of information, and adherence to privacy laws are paramount. For example, unauthorized access to patient data due to jailbreaking not only violates privacy regulations such as HIPAA but can also lead to breaches of trust between patients and healthcare providers. Furthermore, if clinical decision-support tools based on medical LLMs are manipulated through jailbreaking, they could provide incorrect or dangerous recommendations, endangering patient lives.

Table 1 provides a taxonomy of jailbreaking and related security vulnerabilities in medical LLMs, outlining the various ways in which these vulnerabilities can manifest. It is important to note that while this taxonomy aims to be as comprehensive as possible, it is by no means exclusive. The complexities of AI in healthcare create numerous potential points of exploitation, and new vulnerabilities may emerge as both AI technology and the methods used to attack these systems evolve. This taxonomy, therefore, should be seen as an evolving framework that categorizes the most prominent risks, but it must remain adaptable to the dynamic nature of AI in medical contexts.

\begin{table}[ht]
    \centering
    
    \label{table:jailbreaking}
    \begin{tabular}{p{3.75cm} p{4.5cm} p{4.5cm}}
        \toprule
        \textbf{Vulnerability Type} & \textbf{Description} & \textbf{Example} \\
        \midrule
        \textbf{Prompt Injection Attacks} & External users manipulate the input prompt to bypass built-in safeguards, causing the LLM to generate unintended or unsafe outputs. & A malicious user tricks the model into revealing confidential patient data by crafting a prompt that manipulates the model's privacy filters. \\
        \addlinespace
        \textbf{Content Manipulation and Policy Evasion} & Attackers rephrase or subtly alter prompts to evade content moderation filters designed to block inappropriate or harmful content. & A user rephrases a question about illegal drug use, causing the model to provide detailed, unrestricted information on illicit drugs. \\
        \addlinespace
        \textbf{Confidentiality Breaches} & Jailbreaking results in the LLM revealing confidential or private information, such as patient data or proprietary medical information. & The model is manipulated into accessing and revealing sensitive patient information, violating privacy laws like HIPAA. \\
        \addlinespace
        \textbf{Model Misuse for Malicious Intent} & Jailbreaking allows malicious actors to exploit the model to spread misinformation, disinformation, or harmful medical advice. & The model is manipulated to propagate false information about vaccines, leading to potential public health risks and undermining trust in medical authorities. \\
        \addlinespace
        \textbf{Undermining Decision-Support Systems} & Jailbreaking interferes with decision-support functionalities, allowing attackers to manipulate clinical recommendations or alter diagnoses. & An attacker manipulates the model to suggest incorrect drug dosages or misdiagnose conditions, directly compromising patient safety. \\
        \bottomrule
    \end{tabular}
    \caption{Taxonomy of Jailbreaking and Security Vulnerabilities in Medical LLMs}
\end{table}

\subsection{Hallucination}

Hallucinations in medical LLMs can have particularly dangerous consequences in healthcare contexts, where accuracy and trust are paramount. Hallucinations occur when the model generates incorrect, misleading, or fabricated information, which can undermine clinical decision-making, erode trust in AI systems, and ultimately jeopardize patient safety. These hallucinations can manifest in various ways, ranging from factual inaccuracies to logical inconsistencies in diagnostic reasoning. While these manifestations can be categorized into distinct types based on their nature and impact (as shown in Table 2), it is important to recognize that significant overlap often exists between categories.

For instance, a factual hallucination, where the model provides incorrect information (e.g., recommending a non-existent drug), can easily lead to a logical hallucination, where the model’s reasoning becomes flawed due to the faulty data. Similarly, diagnostic hallucinations, where incorrect diagnoses are made, often arise from a combination of factual inaccuracies, semantic misunderstandings, and errors in clinical reasoning. Hallucinations related to treatment recommendations may combine elements from several categories—incorrect information about a treatment (factual), improper reasoning about symptoms or conditions (logical), and unsafe or unsupported recommendations (therapeutic).

These overlaps blur the boundaries between categories, illustrating the complexity of identifying and addressing hallucinations in medical LLMs. While this taxonomy is useful for understanding the distinct risks posed by hallucinations, it is not exhaustive, and other methods of classification may help refine our understanding of these vulnerabilities. Additionally, as medical AI evolves, new types of hallucinations or combinations of existing ones may emerge, necessitating ongoing evaluation of classification frameworks. With this in mind, Table 2 presents an overview of the most prominent hallucination types observed in medical LLMs, but this should be seen as part of a larger, evolving conversation on improving AI accuracy in healthcare.

\begin{table}[ht]
    \centering
    
    \label{table:hallucinations}
    \begin{tabular}{p{3.75cm} p{4cm} p{4cm}}
        \toprule
        \textbf{Hallucination Type} & \textbf{Description} & \textbf{Example} \\
        \midrule
        \textbf{Factual Hallucinations} & The model generates false or fabricated medical information that contradicts established facts. & Suggesting the use of a nonexistent drug like "hydromethrin" for treating hypertension. \\
        \addlinespace
        \textbf{Logical or Reasoning Hallucinations} & The model draws conclusions or recommendations that do not logically follow from the input or violate clinical reasoning. & Recommending chemotherapy for common flu symptoms, which is medically inappropriate. \\
        \addlinespace
        \textbf{Extrinsic Hallucinations} & The model fabricates external references, such as non-existent research studies or clinical trials. & Citing a fictional clinical trial as evidence for a treatment recommendation. \\
        \addlinespace
        \textbf{Intrinsic Hallucinations} & The model generates responses that are unrelated to the input or diverge from the context of the conversation. & Discussing a skin condition when the inquiry was focused on managing diabetes. \\
        \addlinespace
        \textbf{Diagnostic Hallucinations} & The model provides incorrect diagnoses or misinterprets patient symptoms and clinical data. & Diagnosing viral flu symptoms as an autoimmune disorder without proper clinical evidence. \\
        \addlinespace
        \textbf{Semantic Hallucinations} & The model misinterprets or incorrectly uses medical terms, leading to confusion or errors in communication. & Confusing "benign" and "malignant" in the context of a cancer diagnosis. \\
        \addlinespace
        \textbf{Therapeutic Hallucinations} & The model suggests unsafe or unsupported medical treatments that are not aligned with clinical guidelines. & Recommending a herbal remedy for treating a critical condition like heart failure, which has no clinical backing. \\
        \bottomrule
    \end{tabular}
    \caption{Taxonomy of Hallucinations in Medical Language Models}
\end{table}

\section{Guardrails}

The challenges posed by hallucinations, bias propagation, and misinformation in medical LLMs necessitate robust solutions that ensure safety, reliability, and fairness. Guardrails—automated safety mechanisms that enforce model behavior and alignment—offer a promising approach to mitigate these risks. This section introduces popular guardrails and their capabilities.

\subsection{NVIDIA NeMo Guardrails}
NVIDIA NeMo Guardrails \cite{Rebedea2023} use Colang \cite{Colang2024}, an executable programming language, to define clear ethical and operational constraints for LLMs. NeMo Guardrails convert user prompts into vector representations using a K-Nearest Neighbors (KNN) algorithm, comparing them against pre-stored canonical forms. By using semantic similarity functions instead of traditional neural network embeddings, NeMo Guardrails can better capture the meaning of user inputs. The tool then maps these inputs into a multi-dimensional vector space, identifying the most relevant predefined examples to guide the conversation flow.
\begin{figure}[h!]
  \centering
  \includegraphics[width=0.8\textwidth]{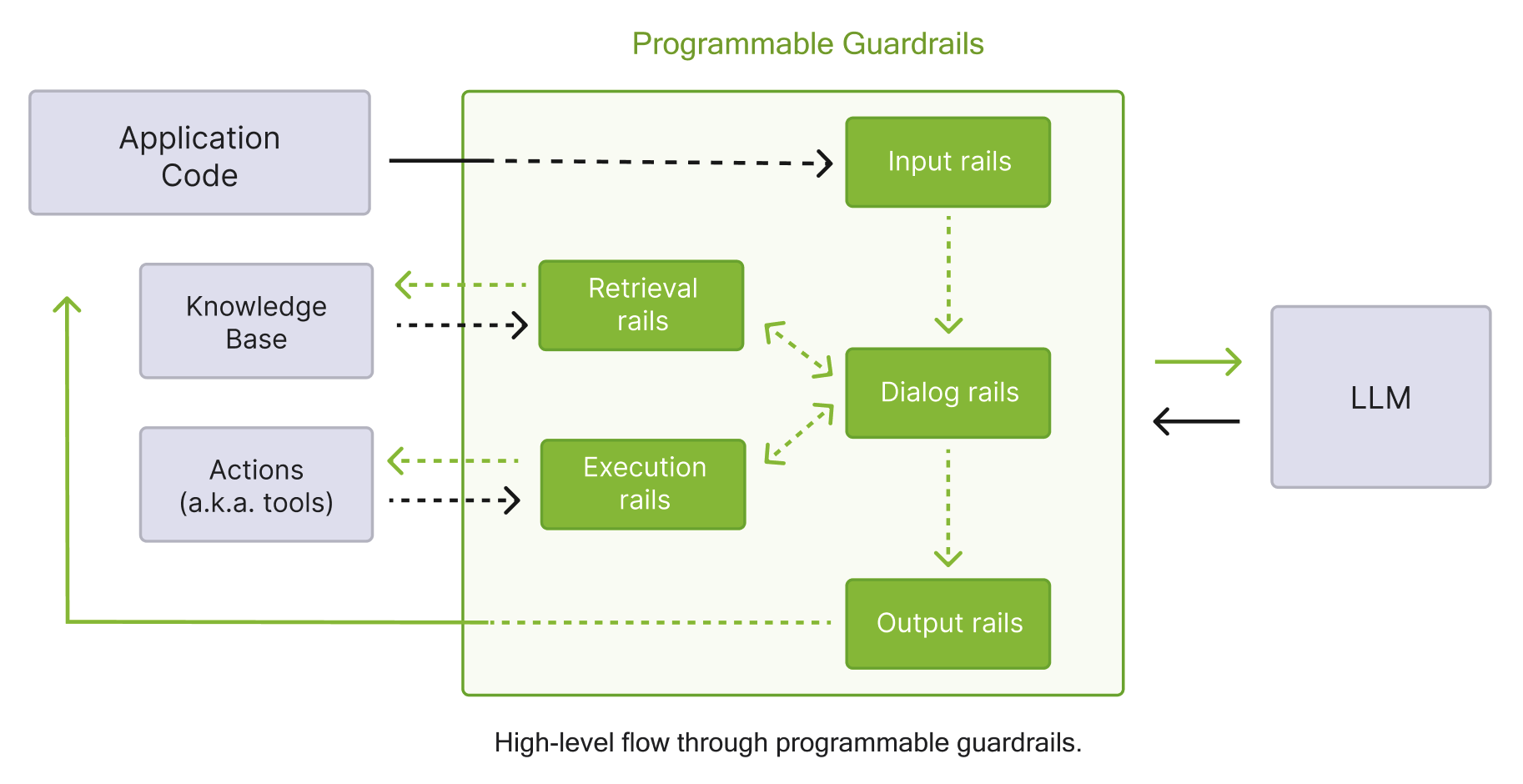}
  \caption{High-level flow through programmable guardrails.}
  \label{fig:guardrails}
\end{figure}

\subsection{Guardrails AI}
Guardrails AI uses RAIL (Reliable AI Markup Language), an XML-based framework, to enforce structure, type, and quality criteria for LLM outputs. RAIL defines expected formats (e.g., JSON), field types (e.g., string, integer), and validation rules (e.g., bias-free text or bug-free code). If outputs fail to meet these standards, corrective actions like reasking or filtering are applied automatically.

\begin{figure}[h!]
  \centering
  \includegraphics[width=0.8\textwidth]{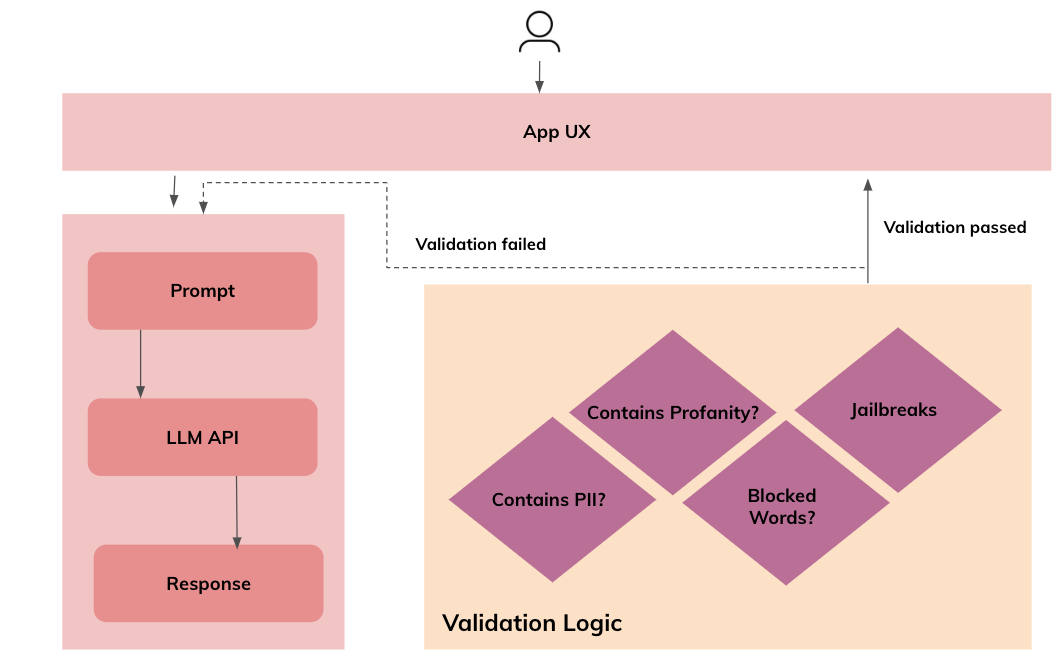}
  \caption{High-level flow of Guardrails AI}
  \label{fig:guardrails}
\end{figure}
RAIL allows for complex, nested structures and ensures precise control over LLM responses, making it ideal for fields requiring strict compliance and accuracy. By wrapping LLM API calls in a Guard object, Guardrails.ai validates and corrects outputs in real-time. Though focused on text, it is designed for future multimodal support.

\subsection{Llama Guard}
Llama Guard is an LLM-based input-output safeguard designed to moderate conversational AI, focusing on classifying safety risks in both user prompts and AI responses. It leverages a safety taxonomy to identify potential risks, including violence, hate speech, and misinformation. By using instruction fine-tuning, Llama Guard enables customizable safety protocols and supports zero-shot and few-shot prompting, allowing adaptation to different taxonomies and guidelines.

Llama Guard performs real-time prompt and response classification, detecting unsafe content and enforcing compliance with predefined guidelines. It uses binary and multi-label classification to assess content safety, offering adaptability across various conversational AI applications. The model is fine-tuned on Meta's Llama3-7b and demonstrates strong performance in safety moderation benchmarks, such as OpenAI Moderation \cite{openai2024moderation} and ToxicChat datasets\cite{lin2024toxicchat}. 

\subsection{Other Guardrails}
OpenAI's GPT-4 utilizes a Content Moderation API \cite{openai2024moderations} to filter and prevent harmful content in LLM outputs. The system detects and flags content that violates categories such as hate speech, violence, self-harm, sexual content (with special emphasis on minors), and harassment. It uses binary labels and probability scores to monitor these categories, ensuring that unsafe content is blocked from reaching users through predefined thresholds.

Claude, developed by Anthropic, employs a Constitutional AI approach \cite{anthropic2023constitutional}for content filtering. This system operates under a set of ethical principles that guide the model during both training and usage, helping it avoid generating harmful outputs. Claude can self-regulate by referring to these ethical rules, which helps in preventing harassment, misinformation, and biased or offensive language, thus minimizing the need for human moderation.

Google's Gemini models incorporate content filtering mechanisms similar to those used in Google’s search products. These systems focus on preventing the generation of hate speech, misinformation, privacy violations, and harmful stereotypes. Gemini's approach is designed to ensure transparency and alignment with responsible AI guidelines, while effectively flagging and filtering potentially harmful content.

Azure OpenAI provides a Content Safety API that detects and filters harmful content in AI-generated outputs. It categorizes content into four key areas: hate speech, violence, sexual content, and self-harm. The system assigns severity scores on a scale from 0 to 6, allowing users to define thresholds for filtering and enforcing content safety. This customizable approach ensures that content meets ethical and safety standards across diverse applications.

\subsection{Comparative Assessment of Guardrail Frameworks}

Each of the guardrails discussed in other sections offers different capabilities in addressing hallucinations, misinformation, and other safety concerns in general domains. The table provides a high-level comparative analysis \cite{shen2024llm}of these guardrails and their effectiveness in various safety aspects.

\begin{table}[ht]
\centering
\begin{tabular}{lccc}
\toprule
\textbf{Abilities}     & \textbf{Llama Guard} & \textbf{Nvidia NeMo} & \textbf{Guardrails AI} \\
\midrule
\textbf{Hallucination}     & \checkmark          & \checkmark           & \checkmark             \\
\textbf{Fairness}          & \checkmark          & -                    & \checkmark             \\
\textbf{Privacy}           & \checkmark          & \checkmark           & -                      \\
\textbf{Robustness}        & -                   & \checkmark           & \checkmark             \\
\textbf{Toxicity}          & \checkmark          & -                    & \checkmark             \\
\textbf{Legality}          & \checkmark          & \checkmark           & -                      \\
\textbf{Out-of-Distribution} & -                  & \checkmark           & \checkmark             \\
\textbf{Uncertainty}       & -                   & -                    & \checkmark             \\
\bottomrule
\end{tabular}
\caption{Abilities among different Guardrails.}
\end{table}

\section{Approach}
To enhance the safety and reliability of language model outputs in medical applications, I have developed a framework that integrates Llama Guard 3, NVIDIA’s NeMo Guardrails, and a fine-tuned medical domain language model, L2M3 \cite{medm24}. While this system can work with any language model, L2M3 was specifically chosen due to its fine-tuning on medical dialogue datasets. In this setup, input validation and jailbreaking detection are managed by Llama Guard 3, followed by processing through NeMo Guardrails for further safety checks. The final response is generated by L2M3, ensuring it is tailored to medical contexts.

The framework receives user inputs which gets passed to Llama Guard 3  to ensure they are sanitized and free from attempts at jailbreaking or violating content policies. By handling input validation at this initial stage, Llama Guard 3 prevents malicious or unintended inputs from affecting subsequent processing and maintains the integrity of the system.

Once the inputs are validated, they are passed to NeMo Guardrails, which enforces additional safety protocols and integrates retrieval rails to access relevant medical information from trusted databases such as the FDA drug database and PubMed. This retrieval provides the necessary context and authoritative information to inform accurate response generation.

The processed input, along with the retrieved medical information, is then fed into L2M3, a language model fine-tuned for the medical domain. L2M3 generates the initial response by leveraging its specialized training to produce accurate and contextually appropriate outputs that align with medical best practices and terminology.

The generated response undergoes further validation against the medical knowledge base to ensure its accuracy. This involves extracting medical terms and drug names using natural language processing techniques and cross-referencing them with the knowledge base to confirm their validity. If any unrecognized or invalid terms are detected, the system prompts L2M3 to correct the errors, providing explicit instructions to verify and amend the invalid terms. The regenerated response is then subjected to the same validation process to ensure all information is accurate and reliable.

Additional safety validations are performed by NeMo Guardrails to ensure the response adheres to ethical guidelines and regulatory standards, including checks for disallowed content and adherence to privacy laws. Once the response passes all validations, it is delivered to the user, maintaining transparency without exposing internal validation mechanisms or compromising user experience.

\subsection{Validation Approach}

To evaluate the effectiveness of the integrated framework, I utilized the Med-HALT \cite{umapathi2023medhalt} dataset. Med-HALT is specifically designed to assess hallucinations in large language models within the medical domain. It offers a set of test cases derived from medical examinations across various countries and specialties, including both reasoning-based and memory-based tests. 

Additionally, I generated a synthetic dataset using Nvidia Nemotron \cite{nvidia2024nemotron4}to further evaluate the system's ability to detect and correct hallucinations. This dataset included intentionally crafted prompts containing fabricated medical terms, such as nonexistent drug names or procedures, to challenge the model's hallucination detection mechanisms. The validation process involved passing these datasets through the integrated system, where Llama Guard 3 handled input validation and hallucination detection, NeMo Guardrails retrieved relevant medical information, and L2M3 generated the responses. By comparing the system's outputs against the annotated ground truths in Med-HALT and the known inaccuracies in the synthetic dataset, I assessed the accuracy, hallucination detection rate, and overall reliability of the framework. 

\section{Results \& Discussion}

To evaluate the effectiveness of integrating Llama Guard 3 and NeMo Guardrails with the L2M3 medical language model, I conducted experiments using two datasets: the Med-HALT dataset and a synthetic dataset generated using Nemotron. The goal was to compare the performance of the L2M3 model (or any other LLMs) without guardrails (baseline) and with guardrails (enhanced) in terms of accuracy, and hallucination detection.

\subsection{Med-HALT Evaluation}
For the validation of the L2M3 medical language model with and without guardrails, I utilized the Med-HALT dataset, which features \textit{Reasoning Hallucination Tests (RHT)} and \textit{Memory Hallucination Tests (MHT)}. The evaluation focused on reasoning-based tests due to their relevance in assessing the model's ability to produce coherent and factually correct outputs without fabricating information.

The Med-HALT tests encompass three distinct types:
\begin{itemize}
\item \textit{False Confidence Testing (FCT)}: Evaluates the model's tendency to provide answers with undue confidence, particularly when lacking adequate information.
\item \textit{Fake Questions Test (FQT)}: Challenges the model with spurious or absurd medical queries to assess its capability to accurately detect and process such inputs.
\item \textit{None of the Above (NOTA) Test}: Determines the model's proficiency in recognizing and disregarding irrelevant or incorrect information.
\end{itemize}

\textbf{Main metrics used are:}

\textit{Accuracy:}

\begin{equation}
\text{Accuracy} = \frac{\text{Number of Correct Predictions}}{\text{Total Number of Predictions}}
\end{equation}

\textit{Final Score} (S):

\begin{equation}
S = \frac{1}{N} \sum_{i=1}^{N} \left[ I\left( y_i = \hat{y}_i \right) \cdot P_c + I\left( y_i \neq \hat{y}_i \right) \cdot P_w \right]
\end{equation}

Where:
\begin{itemize}
    \item \( S \) is the final score,
    \item \( N \) is the total number of samples,
    \item \( y_i \) is the true label of the \( i \)-th sample,
    \item \( \hat{y}_i \) is the predicted label of the \( i \)-th sample,
    \item \( I(\text{condition}) \) is the indicator function that returns 1 if the condition is true and 0 otherwise,
    \item \( P_c \) is the points awarded for a correct prediction,
    \item \( P_w \) is the points deducted for an incorrect prediction.
\end{itemize}

Table 4 highlights the performance improvements achieved by incorporating Guardrails compared to the baseline without Guardrails across all test types in the Med-HALT evaluation.

\begin{table}[htbp]
\centering
\begin{tabular}{lcccccccc}
\toprule
\textbf{Model}                & \multicolumn{2}{c}{\textbf{FCT}} & \multicolumn{2}{c}{\textbf{FQT}} & \multicolumn{2}{c}{\textbf{NOTA}} \\
\cmidrule(lr){2-3} \cmidrule(lr){4-5} \cmidrule(lr){6-7}
                              & \textbf{Accuracy} & \textbf{Score} & \textbf{Accuracy} & \textbf{Score} & \textbf{Accuracy} & \textbf{Score} \\
\midrule
L2M3 Without Guardrails      & 44.38              & 53.14           & 97.26              & 17.58          & 84.11              & 191.6           \\
L2M3 With Guardrails         & \cellcolor[HTML]{D5E8D4}46.12 & \cellcolor[HTML]{D5E8D4}56.08           & \cellcolor[HTML]{D5E8D4}98.11 & \cellcolor[HTML]{D5E8D4}18.0           & \cellcolor[HTML]{D5E8D4}88.0 & \cellcolor[HTML]{D5E8D4}196.0          \\
\bottomrule
\end{tabular}
\caption{\small Performance of L2M3 without and with guardrails on the Med-HALT dataset using FCT, FQT, and NOTA metrics.}

\end{table}

Overall, the integration of guardrails resulted in consistent improvements across all metrics.

\subsection{Synthetic Data Validation}

To further evaluate the effectiveness of guardrails integration, I used a synthetic dataset specifically designed to test hallucination detection and response accuracy. This validation aimed to assess the model's performance using traditional metrics such as accuracy, precision, recall, and F1 score.

The synthetic dataset consists of 16,000 records, divided into two main categories. The first category, Hallucination Instances, comprises 14,000 records, which are further broken down into seven types: \textit{Factual Hallucinations, Logical or Reasoning Hallucinations, Extrinsic Hallucinations, Intrinsic Hallucinations, Diagnostic Hallucinations, Semantic Hallucinations, and Therapeutic Hallucinations}. The second category, Potential Jailbreak Attempts, consists of 2,000 records generated from various sources \cite{SCBSZ24}, specifically designed to evaluate the model's resilience in resisting and effectively managing attempts to bypass guardrails and input validation mechanisms.

\begin{table}[htbp]
\centering
\begin{tabular}{lcccccc}
\toprule
\textbf{Evaluation Aspect}   & \multicolumn{2}{c}{\textbf{Accuracy (\%)}} & \multicolumn{2}{c}{\textbf{Precision (\%)}} & \multicolumn{2}{c}{\textbf{Recall (\%)}} \\
\cmidrule(lr){2-3} \cmidrule(lr){4-5} \cmidrule(lr){6-7}
                             & \textbf{Base} & \textbf{Guardrails}   & \textbf{Base}  & \textbf{Guardrails}  & \textbf{Base}  & \textbf{Guardrails} \\
\midrule
Hallucinations       & 75.0         & \cellcolor[HTML]{D5E8D4}92.8          & 73.5         & \cellcolor[HTML]{D5E8D4}94.5         & 76.0         & \cellcolor[HTML]{D5E8D4}92.1         \\
Jailbreak Attempts           & 68.0         & \cellcolor[HTML]{D5E8D4}96.0          & 66.0         & \cellcolor[HTML]{D5E8D4}96.8         & 70.0         & \cellcolor[HTML]{D5E8D4}94.6         \\
\bottomrule
\end{tabular}
\caption{Performance of L2M3 with Base and Guardrails on the synthetic dataset including hallucinations and potential jailbreaks.}
\end{table}

As highlighted in Table 5, for hallucinations, accuracy increased from 75.0\% to 93.0\%, with the model more effectively identifying and correcting errors to align with medical knowledge. In handling jailbreak attempts, accuracy improved from 68.0\% to 96.0\%, as the guardrails prevented inappropriate content generation and neutralized potential threats, ensuring safer outputs.

\section{Conclusion}
This study highlights the significant improvements in safety, accuracy, and reliability achieved by integrating Llama Guard 3 and NeMo Guardrails with the L2M3 model for medical applications. The guardrails effectively reduce hallucinations, enforce adherence to ethical and regulatory guidelines, and enhance the overall quality of generated responses. As a result, the L2M3 model becomes more robust and dependable, ensuring it meets the stringent requirements of clinical environments where precision and safety are critical.

Furthermore, while this study focuses on the integration with L2M3, the guardrails framework is versatile and can be applied to other language models used in healthcare. This flexibility is particularly important for healthcare applications where incorrect information or unsafe outputs could have serious consequences. By incorporating guardrails into any healthcare-related LLM, the framework ensures the generation of medically accurate and compliant responses, safeguarding against hallucinations and reinforcing the model’s ability to provide safe and reliable information in critical medical scenarios.

In conclusion, this framework offers a vital tool for ensuring the development of AI-driven systems that meet the rigorous safety, accuracy, and ethical standards necessary for healthcare settings.

\medskip
\clearpage

\bibliography{references} 
\bibliographystyle{plain}
\end{document}